\documentclass[aps,prl,reprint,superscriptaddress,showpacs, citeautoscript,floatfix]{revtex4-1}

\usepackage{graphicx}
\usepackage{dcolumn}
\usepackage{bm}
\usepackage{natbib}
\usepackage{amsfonts}
\usepackage{amsmath}
\usepackage{amssymb}
\usepackage{xcolor}

\begin{document}

\title{Enhanced electron-phonon coupling for charge-density-wave formation in La$_{1.8-x}$Eu$_{0.2}$Sr$_{x}$CuO$_{4+\delta}$}

\author{Y.~Y.~Peng}
\email{yingying.peng@pku.edu.cn}\affiliation{Department
of Physics and Materials Research Laboratory, University of
Illinois, Urbana, IL 61801, USA} \affiliation{International Center for Quantum Materials, School of Physics, Peking University, Beijing 100871, China} 
\author{A.~A.~Husain}
\affiliation{Department of Physics and Materials Research Laboratory, University of
Illinois, Urbana, IL 61801, USA} 
\author{M.~Mitrano}
\affiliation{Department of Physics and Materials Research Laboratory, University of
Illinois, Urbana, IL 61801, USA}
\author{S.~Sun}
\affiliation{Department of Physics and Materials Research Laboratory, University of
Illinois, Urbana, IL 61801, USA}
\author{T.~A.~Johnson}
\affiliation{Department of Physics and Materials Research Laboratory, University of
Illinois, Urbana, IL 61801, USA}
\author{A.~V.~Zakrzewski}
\affiliation{Department of Physics and Materials Research Laboratory, University of
Illinois, Urbana, IL 61801, USA}
\author{G.~J.~MacDougall}
\affiliation{Department of Physics and Materials Research Laboratory, University of
Illinois, Urbana, IL 61801, USA}
\author{A.~Barbour}
\affiliation{National Synchrotron Light Source II, Brookhaven National Laboratory, Upton, NY
11973, USA}
\author{I.~Jarrige}
\affiliation{National Synchrotron Light Source II, Brookhaven National Laboratory, Upton, NY
11973, USA}
\author{V.~Bisogni}
\affiliation{National Synchrotron Light Source II, Brookhaven National Laboratory, Upton, NY
11973, USA}
\author{P.~Abbamonte}
\email{abbamonte@mrl.illinois.edu} \affiliation{Department of Physics and Materials Research Laboratory, University of Illinois, Urbana, IL 61801, USA} 

\date{\today}

\begin{abstract}

Charge density wave (CDW) correlations are prevalent in all copper-oxide superconductors. 
While CDWs in conventional metals are driven by coupling between lattice vibrations and electrons, the role of the electron-phonon coupling (EPC) in cuprate CDWs is strongly debated. Using Cu $L_3$ edge resonant inelastic x-ray scattering (RIXS), we study the CDW and Cu-O bond-stretching phonons in the stripe-ordered cuprate La$_{1.8-x}$Eu$_{0.2}$Sr$_{x}$CuO$_{4+\delta}$. We investigate the interplay between charge order and EPC as a function of doping and temperature, and find that the EPC is enhanced in a narrow momentum region around the CDW wave vector. By detuning the incident photon energy from the absorption resonance, we extract an EPC matrix element at the CDW wave vector of $M\simeq$ 0.36 eV, which decreases to $M\simeq$ 0.30 eV at high temperature in the absence of the CDW. Our results suggest a feedback mechanism in which the CDW enhances the EPC which, in turn, further stabilizes the CDW. 

\end{abstract}

\maketitle

Charge density waves (CDWs) are now established as a generic feature of high-$T_c$ cuprate superconductors \cite{KeimerNature}. Evidence for a bulk CDW was first observed in stripe-ordered La-based cuprates near the hole doping $p =1/8$, coinciding with an anomalous suppression of the superconducting critical temperature $T_c$ \cite{TranquadaStripe,FujitaStripe,PeterStripe}. Stripe phases are believed to originate from competition between magnetism and the kinetic energy of charge carriers \cite{Zaanenstripe,Kivelsonstripe}. Since then, CDW phenomena have been discovered in nearly all cuprate families, including Y-, Bi- and Hg-based cuprates \cite{JulienNMR,GiacomoCDW,changCDW,Damascelliscience, PengPRB,COinOD, STMBi2212,HBCOCDW}, where the CDWs are clearly distinct from magnetic order. It is generally accepted that static CDW order competes with superconductivity \cite{GiacomoCDW,changCDW}, though many still believe dynamic CDW order may promote superconductivity \cite{GrilliCDW,dynamicCDW}. The driving force behind charge order in cuprates remains unclear. Several mechanisms have been proposed including, for example, real-space local interactions or Fermi surface nesting \cite{Zaanenstripe,Damascelliscience}, but consensus on the mechanism has not been reached. 

The electron-phonon coupling (EPC) plays an important role in the formation of CDWs in conventional metals \cite{CDWorigin,CDWNbSe2}. In cuprates, inelastic neutron scattering (INS) and non-resonant inelastic x-ray scattering (IXS) studies have revealed strong and ubiquitous renormalization of high-energy Cu-O bond-stretching (BS) phonons and low-energy acoustic phonons around the CDW wave vector \cite{INSphonon1,INSphonon2,IXSphonon1,IXSphonon2}. These observations suggest that the EPC could be strong enough to contribute to the electronic instabilities in the cuprates. It has been suggested that, in the case of stripe-ordered cuprates, the strongest coupling is with the longitudinal BS mode, since its polarization matches the lattice deformation induced by charge stripes \cite{INSphonon2}. This idea motivated us to study whether the electronic coupling with BS phonons can provide insight into CDW formation. Although there have been previous reports of phonon anomalies in stripe-ordered La-based cuprates \cite{INSphonon1,INSphonon2,IXSphonon2}, no study has directly investigated the EPC strength at the CDW wave vector, $Q_{CDW}$. 

With an improved energy resolution, RIXS has recently become an effective tool for directly determining the momentum dependence of the electron-phonon coupling strength \cite{Amentphonon,Tomphonon}. The EPC strength can be extracted from the intensity of the single phonon excitation \cite{Amentphonon} or the ratio between the phonon intensity and its first overtone in RIXS spectra \cite{Amentphonon,LeeCuprate,GrioniTiO2}. However, the former requires a measurement of the absolute intensity, and it is very difficult to experimentally detect the phonon satellites for the latter. A novel method has recently been introduced in which the EPC is estimated by measuring the phonon intensity as the incident beam is detuned from the resonance, reporting an EPC matrix element of $\sim$ 0.17 eV for the BS mode in the insulating parent compound NdBa$_2$Cu$_3$O$_6$ (NBCO) at the Brillouin zone boundary \cite{RossiPRL,Luciophonon}. 

In this Letter, we present a systematic RIXS study of the CDW and Cu-O BS phonons in La$_{1.8-x}$Eu$_{0.2}$Sr$_{x}$CuO$_{4+\delta}$ (LESCO) compounds, which were grown using the traveling solvent floating zone method. We chose LESCO because its low-temperature tetragonal (LTT) structural transition ($\sim$125K) is decoupled from the CDW transition ($T_{CDW}\sim$ 80K), allowing us to isolate the effects relevant to the CDW. A previous resonant x-ray scattering (RXS) experiment reported that there is no CDW signal at \emph{p} $\simeq 0.07$, while the CDW signal is the strongest at \emph{p} $\simeq0.125$ \cite{LESCOphasediagram}. We therefore investigated these two dopings, the aim being to compare them to determine the effect of the CDW on the EPC and {\it vice versa}. Our doping- and temperature-dependent results demonstrate that the EPC is enhanced around the CDW wave vector. Taking the approach of detuning the incident RIXS beam energy, we estimate a coupling constant of $M \simeq$ 0.36 eV for the BS mode in LESCO (\emph{p}$\simeq0.125$) at $Q_{CDW}$ in the CDW phase. In the high temperature normal state, $M$ reduces to $\simeq0.30$ eV. These observations suggest that EPC is important for the CDW formation, and exhibits a feedback phenomenon in which the CDW, in turn, enhances the EPC. 

The RIXS measurements were performed using the new soft x-ray inelastic (SIX) spectrometer at beamline 2-ID at the National Synchrotron Light Source II, Brookhaven National Laboratory \cite{SIXspectrometer}. The RIXS spectra were collected with $\sigma$-incident polarization (perpendicular to the scattering plane) to maximize the charge signal \cite{GiacomoCDW}. The RIXS data, except the detuning spectra, were normalized by the spectral weight of the {\it d-d} excitations following previous conventions \cite{LucioPRB}. The scattering angle was fixed at $130^\circ$ and the incident photon energy was tuned to the resonance maximum of the Cu $L_3$ absorption peak around 931 eV. The total experimental energy resolution was 55 meV. We define our reciprocal lattice units (r.l.u.) in terms of the low-temperature tetragonal unit cell with $a=b=3.79$~{\AA} and $c=13.14$~{\AA}.

\begin{figure}[tbp]
\begin{center}
\includegraphics[width=\columnwidth,angle=0]{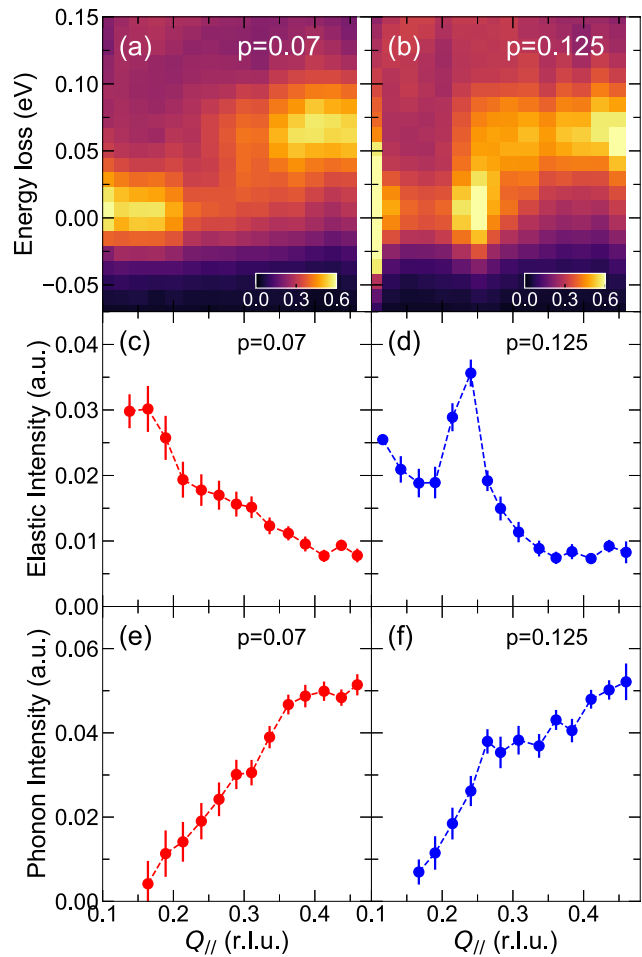}
\end{center}
\caption{Energy/momentum intensity maps of RIXS spectra of La$_{1.8-x}$Eu$_{0.2}$Sr$_{x}$CuO$_{4+\delta}$ along the (0,0)-(0.5,0) symmetry direction taken with  $\sigma$-polarized incident light at 30 K for (a) \emph{p}$\simeq$0.07 and (b) \emph{p}$\simeq$0.125. Integral intensity of the elastic peaks for (c) \emph{p}$\simeq$0.07 and (d) \emph{p}$\simeq$0.125. Integral intensity of the phonon peaks for (e) \emph{p}$\simeq$0.07 and (f) \emph{p}$\simeq$0.125.
\label{Fig1}}
\end{figure}

Figure~\ref{Fig1} displays the momentum-dependent RIXS spectra along the (0,0)-(0.5,0) symmetry direction for \emph{p}$\simeq$0.07 (a) and \emph{p}$\simeq$0.125 (b). 
Two spectral features are visible, a quasielastic line centered at zero energy, which reveals the CDW intensity, and a phonon at $\sim65$ meV whose energy matches the Cu-O BS mode \cite{INSphonon1,Tomphonon}. The intensity of both features is a strong function of the momentum transfer, $Q_\parallel$. The RIXS spectra were fitted using a Gaussian for the elastic peak, a Lorentzian for the phonon peak and an anti-symmetrized Lorentzian for the magnetic excitations, all convolved with a Gaussian energy resolution function. The  resulting integrated quasielastic and phonon intensities are summarized in (Figs. 1(c)-(f)). The strong intensity at small momentum transfer is due to the tails of the elastic peak arising from the specular reflectance from the sample surface. The quasielastic intensity for \emph{p}$\simeq$0.07 does not show a CDW peak (Fig.~\ref{Fig1}(c)), in agreement with previous RXS results \cite{LESCOphasediagram}. On the other hand, the momentum distribution of integrated elastic intensity for \emph{p}$\simeq$0.125 exhibits a pronounced charge order signal at momentum $Q_{CDW}$=0.24 r.l.u. (Fig.~\ref{Fig1}(d)), with a wave vector similar to previous reports from RXS \cite{LESCOvector}. 

\begin{figure}[tbp]
\begin{center}
\includegraphics[width=\columnwidth,angle=0]{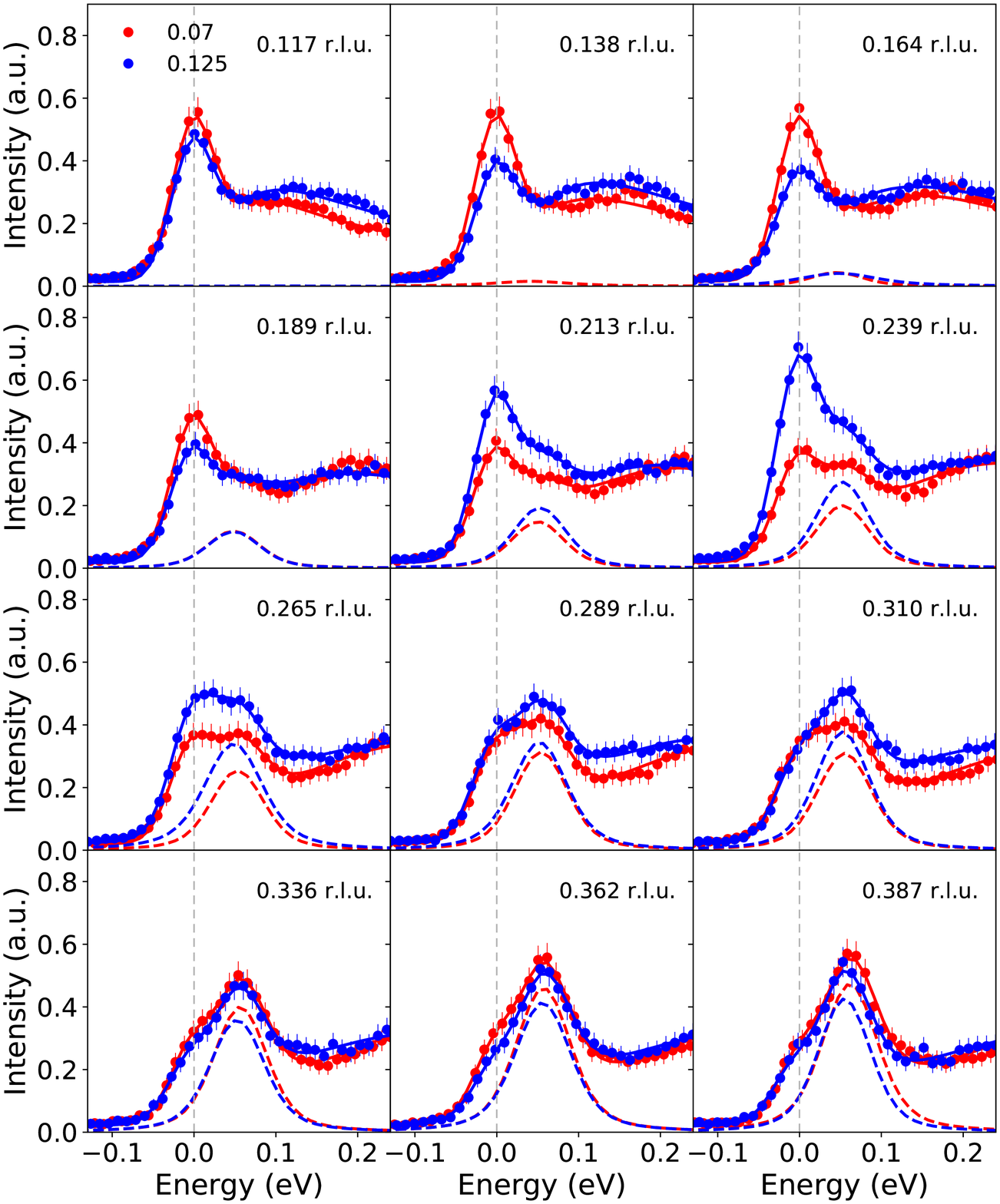}
\end{center}
\caption{Raw energy-loss spectra (markers) and the corresponding fits (solid lines) of RIXS data taken at T=30 K for \emph{p}$\simeq$0.07 (red) and \emph{p}$\simeq$0.125 (blue). The phonon peak fits
are displayed for comparison, indicated by red and blue for \emph{p}$\simeq$0.07 and \emph{p}$\simeq$0.125, respectively.
\label{Fig2}}
\end{figure}

The intensity of the phonon at 65 meV for \emph{p}$\simeq$0.07 increases monotonically with momentum (Fig. 1 (e)), consistent with the prior measurement of Cu-O BS in NBCO \cite{RossiPRL}. By contrast, for \emph{p}$\simeq$0.125 (Fig. 1(f)), the phonon intensity is enhanced for momenta near the CDW momentum. This effect is even clearer in the individual spectra, which are shown in Figure~\ref{Fig2}, comparing the two dopings \emph{p}$\simeq$0.07 and \emph{p}$\simeq$0.125. Due to the limitation of energy-resolution and strong elastic peak at small momentum transfer, we were able to detect unambiguously the phonon signal starting from $Q_\parallel$$\sim$0.14 r.l.u.. At small momenta the phonon peak intensities are comparable between the two dopings. Starting from $Q_\parallel$=0.213 r.l.u., the phonon intensity is higher at \emph{p}$\simeq$0.125 than \emph{p}$\simeq$0.07 until $Q_\parallel$=0.31 r.l.u., after which the phonon intensities of the two become comparable again. Since the phonon intensity in RIXS spectra directly reflects the EPC strength \cite{Amentphonon}, this provides evidence for an enhanced EPC confined to a narrow window of momentum space in the vicinity of $Q_{CDW}$. This agrees with the nonmonotonic phonon intensity as a function of doping in NBCO, which mimics the doping dependence of the CDW signal \cite{Luciophonon}.

\begin{figure}[htbp]
\includegraphics[width=\columnwidth,angle=0]{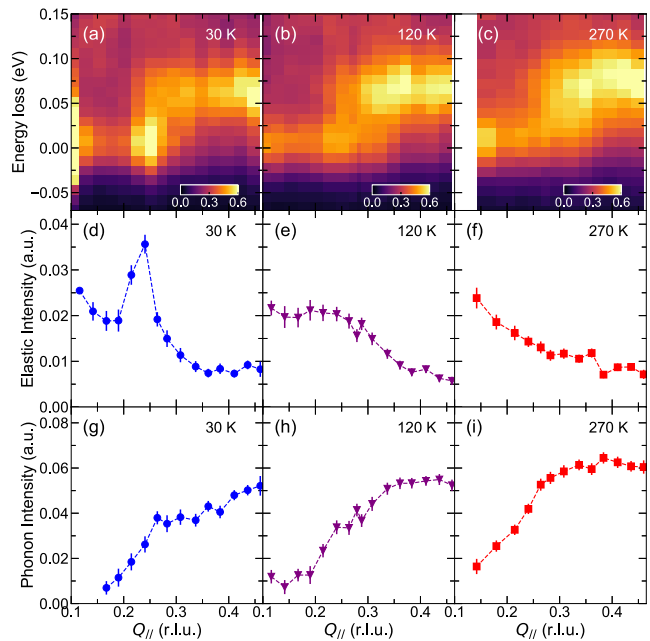}
\caption{Energy/momentum intensity maps of RIXS spectra for \emph{p}$\simeq$0.125 along the (0,0)-(0.5,0) symmetry direction at (a) 30K, (b) 120K and (c) 270K. Integral intensity of the elastic peaks at (d) 30K, (e) 120K and (f) 270K. Integral intensity of the phonon peaks at (g) 30K, (h) 120K and (i) 270K. 
\label{Fig3}}
\end{figure}

To illustrate how this effect changes with temperature, we have conducted the temperature-dependent measurements on \emph{p}$\simeq$0.125, with results shown in Fig.~\ref{Fig3}. The CDW transition temperature, $T_{CDW}$, for this doping is reported to be around 80K \cite{LESCOphasediagram}. We performed measurements at 30K, 120K and 270K across $T_{CDW}$ (Fig.~\ref{Fig3}(a)-(c)). The momentum-dependence of the integrated elastic-peak intensity at 30 K, reproduced from Fig. 1 for convenience, again shows a CDW peak with a full width at half maximum (FWHM) of 0.04 r.l.u.. However, we still observe a broad quasielastic peak around $Q_{CDW}$ at 120 K, which is above $T_{CDW}$, the FWHM increasing to 0.14 r.l.u., as shown in Fig.~\ref{Fig3}(e). This is probably due to dynamic charge fluctuations, reported as the high temperature precursor CDW in LBCO \cite{MiaoPNAS} and the charge density fluctuation in NBCO \cite{GiacomoCDF}. We did not observe any CDW signal when the temperature was raised to 270 K (Fig.~\ref{Fig3}(f)). Correspondingly, we find that the enhancement of phonon intensity near $Q_{CDW}$ at 30 K is reduced at 120 K (Fig.~\ref{Fig3}(g,h)). At 270 K, when the CDW is absent, this enhancement is absent and the phonon intensity again becomes a monotonic function of momentum (Fig.~\ref{Fig3}(i)).

\begin{figure*}[tbp]
\begin{center}
\includegraphics[width=2\columnwidth,angle=0]{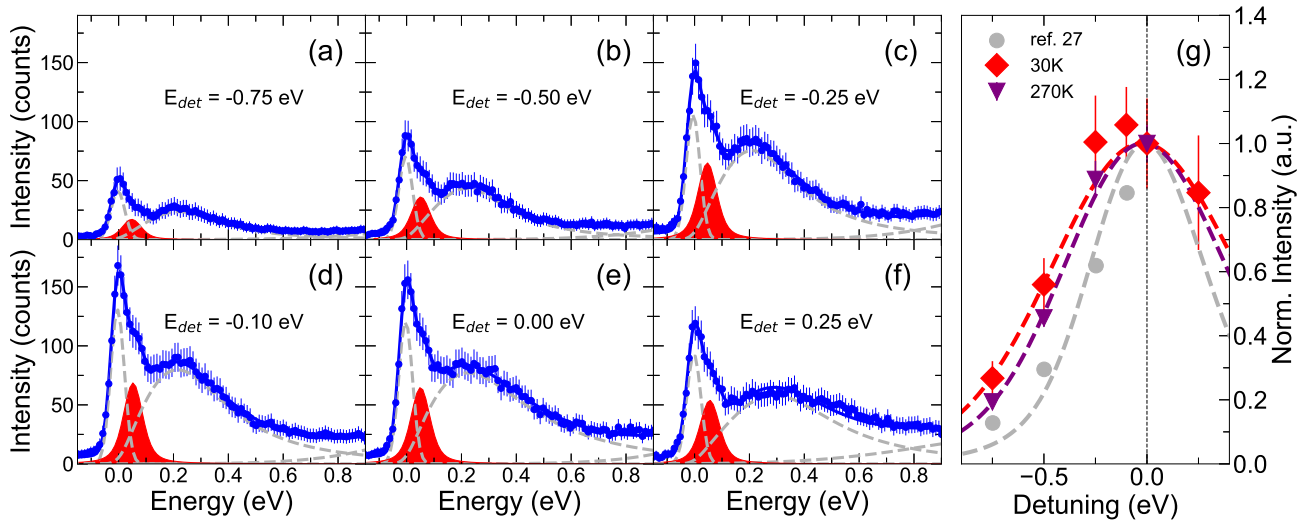}
\end{center}
\caption{(a-f) RIXS spectra of LESCO (\emph{p}$\simeq$0.125) (markers) and the corresponding fits (dashed lines) measured at $Q_{CDW}$= 0.24 r.l.u. taken at \emph{T}=30 K as a function of the detuning energy ($E_{det}$) as indicated. The phonon peak fits are highlighted with filled areas. (g) Detuning dependence of the BS mode at 30K (red diamonds), 270K (violet triangles) and the expected phonon intensity dampening calculated from Eq. 1 (dashed lines). The data from NBCO at $Q_{//}$=0.4 r.l.u. (grey circles) is also displayed for comparison \cite{RossiPRL}.
\label{Fig4}}
\end{figure*}

To estimate quantitatively the EPC strength, we apply the detuning method recently introduced in NBCO  \cite{RossiPRL,Luciophonon}. Using this method it is possible to determine the EPC from a suitable set of spectra measured as a function of detuning energy. As in Ref.~\cite{RossiPRL}, we use the following expression for the phonon intensity, which is based on the approximation that the electronic state couples to a single Einstein phonon with an energy $\omega_0$:

 \begin{equation}\label{Eq1}
   I_{ph} \propto \frac{e^{-2g}}{g} \left| \sum_{n=0}^{\infty} \frac{g^n(n-g)}{n!(\Omega +i\Gamma+(g-n)\omega_0)} \right|^2.
   \end{equation}

\noindent
Here, $g = (M/\omega_0)^2$  is the dimensionless EPC strength at the $Q$ vector of interest, and \emph{M} is the absolute value of the EPC matrix element. $\Gamma$ is the intrinsic width of the Cu $L_3$ resonance, which we take to be 0.28 eV \cite{Coreholelife}, and the phonon energy $\omega_0$=0.06 eV as determined from the RIXS data. The detuning energy $\Omega$ = $\omega-\omega_{res}$, where $\omega$ is the incident photon energy and $\omega_{res}$ is the resonant maximum of the x-ray absorption spectrum. We normalize the detuning curve to the value at this maximum and then fit the measured detuning curve by using Eq. 1 to obtain the values of \emph{g} and $M$. 

Figure \ref{Fig4}(a)-(f) shows the RIXS spectra for \emph{p}$\simeq$0.125 collected at $Q_{CDW}$=0.24 r.l.u. at 30K for various detuning energies, $\Omega$. The self-absorption effects have been corrected for these spectra following the procedures in refs. \cite{RossiPRL,MinolaPRL}. The phonon component is highlighted and the phonon spectral weight is plotted as a function of the detuning energy in Fig.~\ref{Fig4}(g). As expected the spectral intensity drops rapidly as energy is detuned. A least-squares fit of the data at 30K using Eq. 1 with a cut-off at \emph{n}=70, beyond which the fitting was unchanged, gives the optimized values of  \emph{g}=(34.2$\pm$5.4) and \emph{M}=(0.35$\pm$0.03) eV. We also performed the detuning measurements at 270 K at the same momentum in the absence of a CDW signal. The phonon spectral weight as a function of detuning energy is displayed in Fig.~\ref{Fig4}(g), with the intensity decreasing faster than detuning at 30 K. The fit values for 270 K are \emph{g}=(24.7$\pm$1.3) and \emph{M}=(0.3$\pm$0.01) eV. Thus, within the theory of Refs. \cite{RossiPRL,Luciophonon}, the EPC is reduced in the absence of the CDW. It is interesting that the EPC strength in LESCO is very close to \emph{M}$\sim$0.34 eV recently found in prototypical CDW compound NbSe$_2$ at $Q_{CDW}$\cite{FengPhonon}. This could imply a common microscopic role of EPC for CDWs in cuprates and transition-metal dichalcogenides. 

Figure~\ref{Fig4}(g) compares the current data to the results of the study of the breathing mode in antiferromagnetic insulating NBCO reported in Ref. \cite{RossiPRL}, in which a CDW was absent. 
The phonon intensities in LESCO (\emph{p}$\simeq$0.125) at both temperatures decay much slower than in NBCO. This is consistent with the robust phonon intensity upon detuning in optimal-doped NBCO which has a weak CDW signal \cite{Luciophonon}. The EPC constants obtained in insulating NBCO are \emph{g} = (6.6$\pm$2.2) and \emph{M}=(0.18$\pm$0.03) eV for the breathing mode. Thus in the stripe-ordered LESCO we obtain \emph{M}$\sim$2 times the value in NBCO and the EPC in LESCO remains large at high temperature. This indicates the EPC is intrinsically larger in LESCO (\emph{p}$\simeq$0.125). It should be stressed that at this doping three transition temperatures $T_{LTT}$=125 K, $T_{CDW}$=80 K and the spin transition $T_{SO}$=45 K are distinctly different. These indicate the formation of CDW is not driven by the structure transition or magnetic order due to the stabilization between antiferromagnetic domains \cite{Zaanenstripe}, though it can be stabilized by the structural distortions \cite{TranquadaStripe} and couples with spin order at low temperatures \cite{CoupleSO,MiaoPNAS}. Our results support a picture in which lattice excitations play a central role in the formation of the CDW, and furthermore exhibit a feedback effect in which the CDW enhances the EPC, which further stabilizes the CDW. 

\begin{acknowledgments}
The authors are grateful to J. van den Brink, G. Ghiringhelli, L. Braicovich, S. Johnston, Y. Li, J. Feng, M. Rossi, T. Devereaux and W. S. Lee for useful discussions. This work was supported by the U.S. Department of Energy, Office of Basic Energy Sciences grant no. DE-FG02-06ER46285. P.A. gratefully acknowledges support from the EPiQS program of the Gordon and Betty Moore Foundation, grant GBMF4542. Crystal growth was supported by DOE grant DE-SC0012368. This research used beamline 2-ID of the National Synchrotron Light Source II, a U.S. Department of Energy (DOE) Office of Science User Facility operated for the DOE Office of Science by Brookhaven National Laboratory under Contract No. DE-SC0012704.
\end{acknowledgments}

\end{document}